\documentstyle[preprint,aps]{revtex}
\begin{document}

\draft

\title{Small and large polarons in nickelates, manganites, and cuprates}
\author{P. Calvani, P. Dore, S. Lupi, A. Paolone, P. Maselli, and P. Giura}
\address{INFM - Dipartimento di Fisica, Universit\`a di Roma ``La Sapienza'',
Piazzale A. Moro 2, I-00185 Roma, Italy}
\author{B. Ruzicka}
\address{Daresbury Laboratories, Warrington WAK 4AD, United Kingdom}
\author{S-W. Cheong} 
\address{Bell Laboratories, Lucent Technologies, Murray Hill, New Jersey 07974,
U.S.A.}    
\author{W. Sadowski}
\address{Faculty of Applied Physics and Mathematics, Technical University of
Gdansk,}
\address{G. Narutowicza 11/12, 80-592 Gdansk, Poland} 
\maketitle

\begin{abstract} 
By comparing the optical conductivities of La$_{1.67}$Sr$_{0.33}$NiO$_{4}$ 
(LSNO), Sr$_{1.5}$La$_{0.5}$MnO$_4$ (SLMO), Nd$_2$CuO$_{4-y}$ (NCO), and 
Nd$_{1.96}$Ce$_{0.04}$CuO$_{4}$ (NCCO), we have identified a
peculiar behavior of polarons in this cuprate family. While in LSNO and SLMO 
small polarons localize into ordered structures below a transition
temperature, in those cuprates the polarons appear to be large, and at low $T$ 
their binding energy decreases. This reflects into an increase of the polaron 
radius, which may trigger coherent transport. 
\end{abstract}

La$_{2-x}$Sr$_x$NiO$_{4+y}$ (LSNO) and Sr$_{2-x}$La$_x$MnO$_4$ (SLMO) 
are isostructural to the high-T$_c$ superconductor La$_{2-x}$Sr$_x$CuO$_{4+y}$
(LSCO). Their parent compounds La$_{2}$NiO$_{4}$  and Sr$_{2}$MnO$_4$ are 
charge-transfer, antiferromagnetic insulators like La$_2$CuO$_4$. Nevertheless,
neither LSNO, nor SLMO exhibit superconductivity. Both of them are even poor 
conductors: 
the former becomes metallic only for $x \sim 1$, in the latter no metallic 
phase has been observed yet. Recently, evidence for the formation of polarons 
(bipolarons) in LSNO (SLMO) and for their ordering below $\sim$ 220 K 
($\sim 250$ K) has been reported.\cite{Chen,Bao}
In the cuprate families which show High-T$_c$ superconductivity, polaron 
formation induced by doping has been detected in several infrared 
experiments.\cite{Bucher,Falck92,prb96,Yagil} However, no charge ordering 
has been observed by diffraction techniques in the superconducting materials,
even if EXAFS data on BSCCO and LSCO have been interpreted in terms of charged 
stripes, possibly short-living.\cite{Bianconiprl}
                      
In the present paper, a comparison is proposed among the infrared polaron bands
in some nickelates, manganites, and cuprates, aimed at detecting any anomalous
behavior in the latter compounds with respect to the other perovskites. Such 
comparison should be particularly meaningful at low temperature, where the 
superconducting transition takes place in cuprates only. 
The optical densities extracted from 
transmission measurements on polycrystalline La$_{1.67}$Sr$_{0.33}$NiO$_{4}$, 
Sr$_{1.5}$La$_{0.5}$MnO$_4$, and Nd$_2$CuO$_{4-y}$ (NCO) dissolved in a
CsI matrix, as well as the optical conductivity obtained from the reflectivity 
of a single crystal of Nd$_{1.96}$Ce$_{0.04}$CuO$_4$ will then be compared 
with each other. 

The experimental procedure has been reported previously.\cite{rc96} 
Once the optical conductivity $\sigma (\omega)$ (or the optical density $O_d$ 
$\propto$ $\sigma (\omega)$, see
Ref. \onlinecite{rc96}) of a doped charge-transfer insulator has been
determined, it can be fitted to the following general expression: 
   
\begin{equation}
\sigma (\omega) = \sigma_{ph} (\omega) + \sigma_{IRAV} (\omega) 
+ \sigma_{pol} (\omega) + \sigma_{MIR} (\omega) + \sigma_{CT} (\omega). 
\label{sigma}
\end{equation}

\noindent
Here $\sigma_{D} (\omega)$ is the Drude conductivity from quasi-free carriers, 
$\sigma_{ph} (\omega)$ is the contribution of the extended TO phonons,
$\sigma_{IRAV} (\omega)$ that of the local modes induced by the self-trapped 
charges which distort the lattice, $\sigma_{MIR} (\omega)$ is the 
$T$-independent part of the midinfrared absorption (possibly due to states 
created by chemical doping in the charge-transfer gap),
and $\sigma_{CT} (\omega)$ is the charge-transfer band. All those terms are 
usually reproduced by Lorentzians multiplied by $4 \pi/\omega$.\cite{prb96} 

Analytical expressions for the contribution $\sigma_{pol} (\omega)$ have been
derived by Reik\cite{Reik} in the case of a small polaron which undergoes 
adiabatic hopping, and by Devreese {\it et al.}\cite{Devreese} and 
Emin\cite{Emin} in the 
case of a large polaron extended over several lattice sites. For a small 
polaron one may write:\cite{Reik} 

\begin{equation}
\sigma_{pol}^{small} (\omega) \propto (n_p/\omega \Delta) 
sinh (4E_p \omega/ \Delta^2) exp[-(\omega^2+4E_p^2)/\Delta^2] 
\label{small}
\end{equation}

\noindent
where $n_p$ is the polaron concentration at $T$, $\omega$ is the photon energy,
$\Delta = 2 \sqrt{2E_pE_{vib}}$, and $E_p$ is the polaron binding energy. 
One may put in Eq.\ (\ref{small}) 
$E_{vib}$ = (1/2) $\hbar \omega^*$ in the low-$T$ limit, where $\omega^*$ is a 
characteristic phonon frequency, and $E_{vib} \sim kT$ in the high-$T$
limit.\cite{Emin} All quantities are expressed in cm$^{-1}$. For large values 
of $E_p/\omega^*$, the maximum absorption will occur at 
$\omega \simeq 2E_p/ \hbar$. Hopping with simultaneuous annihilation (for 
$\omega < 2E_p/ \hbar$) or creation (for $\omega > 2E_p/ \hbar$) of optical 
phonons causes the broadening of the polaron band.      
In turn, the large-polaron conductivity can be written as:\cite{Emin}  

\begin{equation}
\sigma_{pol}^{large} (\omega) \propto {(n_p(T)/\omega) a^2 (\omega - 3E_p) 
\over {[1+ a^2 (\omega - 3E_p)]^3}} 
\label{large}
\end{equation}

\noindent
where $a \propto \sqrt {2mR}$, with $m$=polaron effective mass and 
$R$=polaron radius. One may notice
that the dependence on temperature here comes from $n_p(T)$. Moreover, 
the absorption in Eq.\ (\ref{large}) has a threshold at $3E_p$, even if the 
large polaron may also perturb the far infrared phonon spectrum at frequencies 
much lower than $E_p/\hbar$.\cite{Emin2}                 

The experimental optical density $O_d(\omega)$ is shown in Fig. 1 at low 
(20 K) and high (300 K) temperature for the polycrystalline sample 
La$_{1.67}$Sr$_{0.33}$NiO$_{4}$, a  
powder where a charge/lattice modulation along the diagonals of the 
Ni-O squares on the $a-b$ planes, with a period 3${\sqrt 2} a$, has been
observed below 250 K.\cite{Chen} The spectrum at 300 K in Fig. 1
shows a broad polaronic background superimposed to phonon peaks. It extends 
from the lowest frequencies to the charge-transfer (CT) band, peaked at 10200 
cm$^{-1}$. This polaronic continuum at $\omega < 2E_p$ shows that the initial 
phonon states at high energies are populated. 
Thus, the energy needed for the charge to move from its perturbed site to a 
neighboring, unperturbed, site is
comparable with that of thermal excitations, and the polarons are mobile. 
This picture changes drastically at low $T$, as shown by the 20 K spectrum: 
here, a depopulation of the phonon states opens an energy gap in the
background below $\sim$ 700 cm$^{-1}$, so that a midinfrared band
peaked at $\sim$ 2000 cm$^{-1}$ shows up. Thus, at low $T$ the charges remain 
self-trapped until photons of suitable energy promote intersite hopping. 

The formation of charged superstructures between 250 and 200 K, as reported in 
diffraction experiments, is confirmed by the behavior of the strong phonon band 
around 350 cm$^{-1}$ in the inset of Fig. 1. In our powder, this results 
from a superposition of the twofold $E_{u}$ bending  of the 
in-plane Ni-O bond, and from the $A_{u}$ mode which displaces the Ni atom and 
both apical oxygens along the $c$-axis.\cite{Bates} In the pure nickelate,
the three vibrations should be observed at approximately the same energy. 
In LSNO at 300 K, the charges induce random deformations of the octahedra which
cause an inhomogeneous broadening of the peak. Below 250 K, the ordering
transition induces regular deformations which give rise to the sharp phonon
doublet shown in the inset. 
  
If one fits Eq.\ (\ref{sigma}) to the data of Fig. 1, with 
$\sigma_{pol} (\omega)$ given by the small-polaron Eq.\ (\ref{small}), one
gets the dashed curves plotted in Fig. 1. Except for the very far infrared,
where the transmittance of the CsI matrix is poor,
the absorption spectrum is well reproduced  by the model, where  Lorentz 
oscillators at 3200 cm$^{-1}$ and at 13500 cm$^{-1}$ have been introduced 
for $\sigma_{MIR} (\omega)$ and $\sigma_{CT} (\omega)$, respectively. 
The 300 K fit yields a polaron bandwidth $\Delta$ = 1233 cm$^{-1}$ and 
2$E_p$ = 1945 cm$^{-1}$, in excellent agreement with the high-$T$  
prediction $\Delta = 2\sqrt {2E_p kT}$ which gives 2$E_p$ = 1900 cm$^{-1}$  
The 20 K fit in turn yields 2$E_p$ = 2000 cm$^{-1}$ and, from the low-$T$ 
expression $\Delta = 2\sqrt {E_p \omega^*}$ = 1050 cm$^{-1}$, one
finds $\omega^*$ = 275 cm$^{-1}$. 

If in the perovskite one replaces Ni by Mn, one finds effects similar to those
of Fig. 1, suggestive of an even stronger charge-lattice interaction.
In Fig. 2 one can see the optical density of 
Sr$_{1.5}$La$_{0.5}$MnO$_4$, a sample where commensurate charge/magnetic 
superlattices similar to those of La$_{1.67}$Sr$_{0.33}$NiO$_{4}$ 
have been detected by electron diffraction,\cite{Bao} and by neutron
scattering.\cite{Sternlieb} In SLMO, however, the superlattices are 
attributed to the ordering of small bipolarons.\cite{Bao} At 300 K, 
a deep absorption minimum is already present at 750 cm$^{-1}$, pointing toward 
low polaron mobilities even at room temperature. Here, the broad background 
observed in LSNO at 300 K should then appear at much higher temperatures.
However, the behavior at low $T$ is not qualitatively different, as the minimum
in Fig. 2 further deepens, through {\it a transfer of spectral weight  
towards higher energies}. At $T<$ 100 K a polaron-like peak is left, which is
peaked at a much higher energy than in Fig. 1 and can be attributed to the
bipolaronic superstructure found out below $\sim$ 220 K.\cite{Bao} Such 
a charge ordering is confirmed here again by 
splittings in the intense E$_{u}$-A$_{u}$ manifold (inset of Fig. 2). At room 
temperature, a broad absorption peak with a few shoulders is observed,  
suggestive of disordered octahedra deformations. Between 250 and 200 K, three  
lines appear at 346, 388, and 433 cm$^{-1}$. The detection of a 
triplet suggests that two charges on a single site 
remove both the twofold degeneracy of the Mn-O bending mode and the accidental 
degeneracy between this mode and the A$_{u}$ vibration. 

Let us now consider the infrared spectrum of a cuprate like NCO, 
the parent compound of the e-doped 
high-$T_c$ superconductor Nd$_{2-x}$Ce$_x$CuO$_{4-y}$ (NCCO). The infrared
$a-b$ plane reflectivity of both insulating (NCO) and metallic (NCCO) single 
crystals has been extensively studied, and found to exhibit polaronic 
contributions centered at $\sim$ 1000 cm$^{-1}$.\cite{prb96,Lupi,Thomas92} 
Here, we have measured the infrared transmission 
a Nd$_2$CuO$_{4-y}$ (NCO) pellet prepared in the same way as the perovskites
of Figs. 1 and 2 to facilitate the comparison.  The 
structure of NCO differs slightly from that of LSNO and SLMO, as in the former 
perovskite the apical oxygens are displaced on the
walls of the tetragonal cell. 

The optical density $O_d$ of the NCO pellet, where extra charges are introduced
by oxygen deficiency, is shown in Fig. 3. The farinfrared absorption shows the 
four TO phonon peaks of the $a-b$ plane, which partially hide the three TO 
modes polarized along the $c$-axis. Indeed, the frequencies of these latters 
are very close to those of the first, second, and fourth $a-b$ mode, 
respectively.\cite{Heyen} At low $T$ one may also notice a few shoulders 
on the TO phonons, the IRAV modes induced by doping.\cite{prb96} 
In the midinfrared, where in single crystals one distinguishes two peaks 
centered at $\sim$ 1000 cm$^{-1}$ ($d$ band) and at $\sim$ 3500 cm$^{-1}$ 
(MIR band), respectively,\cite{prb96} Fig. 3 shows a broad absorption 
delimited by a gap below $\sim$ 800 cm$^{-1}$. If one now compares
the optical densities of Fig. 3 with those of the Ni, Mn oxides in Figs. 1 and 
2, one remarks that their temperature behavior is different. Indeed, in the Ni,
Mn oxides the gap between the phonon region and the polaron band {\it deepens}
as $T$ lowers. On the contrary, in the reduced cuprate
the intensity of the background slightly increases at low $T$, 
so that the gap at $\sim$ 800 cm$^{-1}$ tends to be partially filled. In other 
words, the few polaronic charges present in NCO seem to increase their mobility
at low temperature, instead of localizing as observed in LSNO and SLMO. 

In order to verify this unexpected behavior in a more doped sample, and to
check whether it can be ascribed to the $a-b$ plane, we have also studied a 
single crystal of Nd$_{2-x}$Ce$_{x}$CuO$_{4}$. This latter is expected to be 
more heavily doped than NCO, but its Ce concentration ($x$=0.04) is much lower 
than that reported for the insulator-to-metal transition ($x$=0.12). Fig. 4 
shows the $a-b$ plane optical conductivity $\sigma (\omega)$ of the above NCCO 
crystal. The solid lines are extracted by Kramers-Kronig transformations
from the normal-incidence reflectivity $R(\omega)$, as measured from 75 through 
20000 cm$^{-1}$. The latter has been accurately extrapolated to zero frequency 
by fitting the reflectivity to a model Drude-Lorentz dielectric function, then 
by reconstructing the missing part of $R(\omega)$ through the
fitting parameters. At high frequency, the extrapolation has been based on 
existing NCO data up to 40 eV.\cite{Tajima} In Fig. 4, the 
$\sigma (\omega)$ thus obtained exhibits at 300 K the expected 
insulator-like behavior and a $d$ 
band peaked at $\sim$ 1000 cm$^{-1}$. The latter, partially superimposed 
to a broad midinfrared band, can be fitted by the large polaron 
model of Eq.\ (\ref{large}), with the characteristic threshold at $3E_p$, 
here found out at $\sim$ 500 cm$^{-1}$. One may notice that around 
$E_p \simeq$ 170 cm$^{-1}$, the 300 K curve shows a broad absorption that has 
been also observed in previous experiments on the same compound.\cite{prb96} 
All of the four $E_u$ phonons of NCCO exhibit Fano-like lineshapes,\cite{Fano} 
possibly due to interactions with the polaronic background, as it will 
be discussed in detail in a separate paper.\cite{Lupi97} The resulting fit, in 
excellent agreement with the measured $\sigma (\omega)$, is shown by a dashed
line in Fig. 4.

As the temperature is lowered, the spectrum of Fig. 4 changes drastically. 
At 20 K, the $d$ band looses much of its intensity, its peak moves towards 
lower frequencies, while two peaks appear at the lowest frequencies here
measured. The former is the narrow contribution at $\sim$ 100 cm$^{-1}$. The 
latter, which extends to the lowest frequencies here measured, suggests 
the appearance of a strong Drude-like contribution. Both features 
are insensitive to the choice of the extrapolation to zero frequency, as
verified by comparing the results of the above procedure based on a
Drude-Lorentz fit with a standard Hagen-Rubens approximation. 
The fit of the 20 K curve by the large polaron model of Eq.\ (\ref{large})  
places the 3$E_p$ threshold at 300 cm$^{-1}$, so that $E_p$ here coincides with
the sharp peak at 100 cm$^{-1}$. As the large-polaron radius 
is\cite{Alex} $R_p \approx (\alpha^2 m \omega ^*)^{-1} \approx (E_p m)^{-1}$, 
where $m$ is the effective mass of a quasi-free electron, the softening 
here observed between 300 and 20 K 
suggests that $R_p$  increases at low temperature. As one also observes at
the same $T$ the appearance of a Drude-like peak, one may speculate that 
the increase in the polaron size may trigger a coherent polaronic transport. 
The impressive {\it transfer of spectral weight towards lower energies} 
observed in NCCO at low temperature can be better appreciated in the 
inset of Fig. 4, where the difference 
$\Delta \sigma (\omega)$ between the optical conductivities at 20 K and 300 K,
once subtracted of the Lorentzian contributions from phonons and local modes, 
is plotted  vs. $\omega$. One can see how the polaron band at $\sim$ 1000 
cm$^{-1}$ looses weight at low temperature in favor of the peak at $\sim$ 100 
cm$^{-1}$ and of a Drude peak centered at zero frequency. This behavior,
just opposite to that of hopping small polarons 
in the nickelate and  the manganite of Figs. 1 and 2, confirms and extends the 
result of Fig. 3. In the heavily doped cuprate the polarons are large, and at
low temperature their radius increases so that coherent carrier states may 
become available, as shown by the insurgence of a strong Drude term. 
This surprising effect appears to be similar to an anomalous enhancement 
in the Drude optical conductivity  observed at low temperature\cite{Lobo} 
in superconducting La$_2$CuO$_{4.06}$.  

In conclusion, the infrared region of all the perovskitic oxides here examined
is dominated by polaronic effects. Nevertheless, while the optical
conductivities of the nickelate LSNO and of the manganite SLMO are easily 
interpreted in terms of charge localization and ordering at low temperature, 
in excellent agreement with the electron diffraction experiments,
$\sigma(\omega)$ in the low-doping cuprate NCO shows that the fraction of 
mobile charges increases slightly as the temperature decreases. In a Ce-doped 
NCCO single crystal, whose midinfrared absorption is well fitted
by a large-polaron model with a polaron binding energy which decreases at low
temperature, a peak appears in the extreme farinfrared at 20 K suggesting 
the onset of a Drude-like behavior. This result shows that at temperatures
that are typical of the superconducting transition in more doped cuprates, 
the normal-state conductivity of a semiconducting member of the family may 
become metallic due to an increase 
in the average polaron radius. It can be emphasised once again that 
such optical behavior is opposite to that displayed by the small polarons in 
the perovskites of Ni and Mn, which indeed are poor metals and do not exhibit 
superconductivity.

\acknowledgments
We wish to thank M. Capizzi, D. Emin, S. Ciuchi, and G. Iadonisi for valuable 
discussions. This work has been supported in part by the 
HC\&M programme of the European Union under the contract 94-0551.
 

\vfill \eject

                                                           
\begin{figure}
\caption{The optical density $O_d$  of polycrystalline 
La$_{1.67}$Sr$_{0.33}$NiO$_{4}$ is reported at 300 K and 20 K (solid lines), 
and fitted to Eq.\ (\ref{sigma}) by use of the small-polaron conductivity of 
Eq.\ (\ref{small}) (dashed lines). The region of the  E$_{u}$ - A$_{u}$ 
manifold is reported in the inset at five different temperatures (the dotted 
line refers to 300 K, the solid one to 20 K).} 
\label{fig1}
\end{figure} 

\begin{figure}
\caption{The optical density $O_d$ of polycrystalline 
Sr$_{1.5}$La$_{0.5}$MnO$_4$ is reported at 300 K (dashed line) and 20 K (solid 
line). The region of the  E$_{u}$ - A$_{u}$ manifold is reported in the inset 
at five different temperatures (the dotted line refers to 300 K,
the solid one to 20 K).} 
\label{fig2}
\end{figure} 

\begin{figure}
\caption{The optical density $O_d$ of polycrystalline Nd$_2$CuO$_{4-y}$ 
at 300 K (dashed line) and 20 K (solid line).}
\label{fig3}
\end{figure} 

\begin{figure}
\caption{The $a-b$ plane optical conductivity $\sigma(\omega)$ of a single 
crystal of Nd$_{1.96}$Ce$_{0.04}$CuO$_4$ at 300 K and 20 K 
(solid lines) is fitted to Eq.\ (\ref{sigma}), with $\sigma_{pol} (\omega)$ 
given by the large-polaron model of Eq.\ (\ref{large}) (dashed lines). 
In the inset, the difference 
$\Delta \sigma (\omega)$ between the optical conductivities at 20 K and 300 K,
once subtracted of the contributions from phonons and local modes, is plotted 
in units 10$^3 \Omega^{-1}$cm$^{-1}$ vs. $\omega$ in cm$^{-1}$.}
\label{fig4}
\end{figure}


\input epsf  
\epsfbox{fig1-jsup97.epsf}

\input epsf  
\epsfbox{fig2-jsup97.epsf}

\input epsf  
\epsfbox{fig3-jsup97.epsf}

\input epsf  
\epsfbox{fig4-jsup97.epsf}

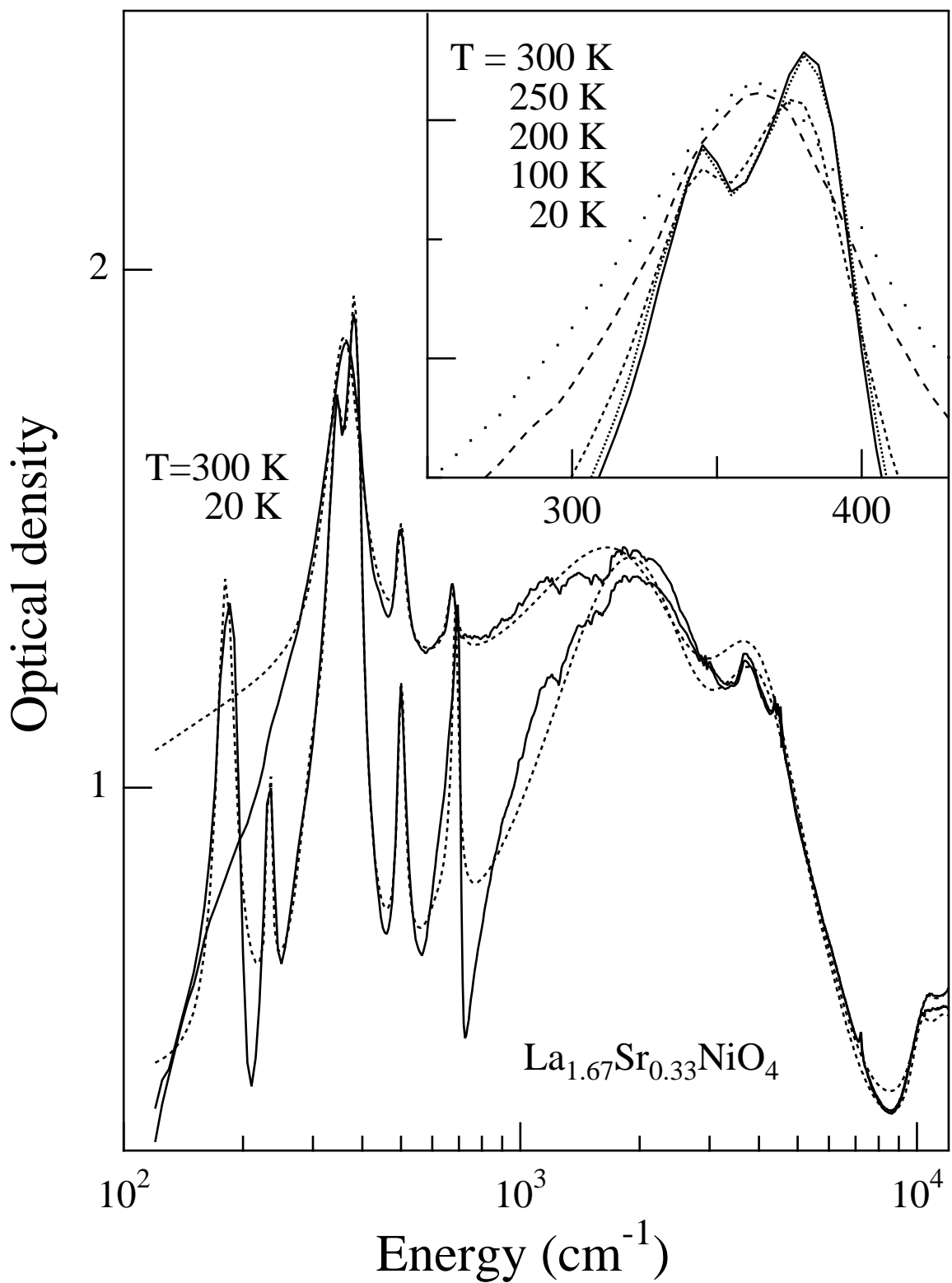
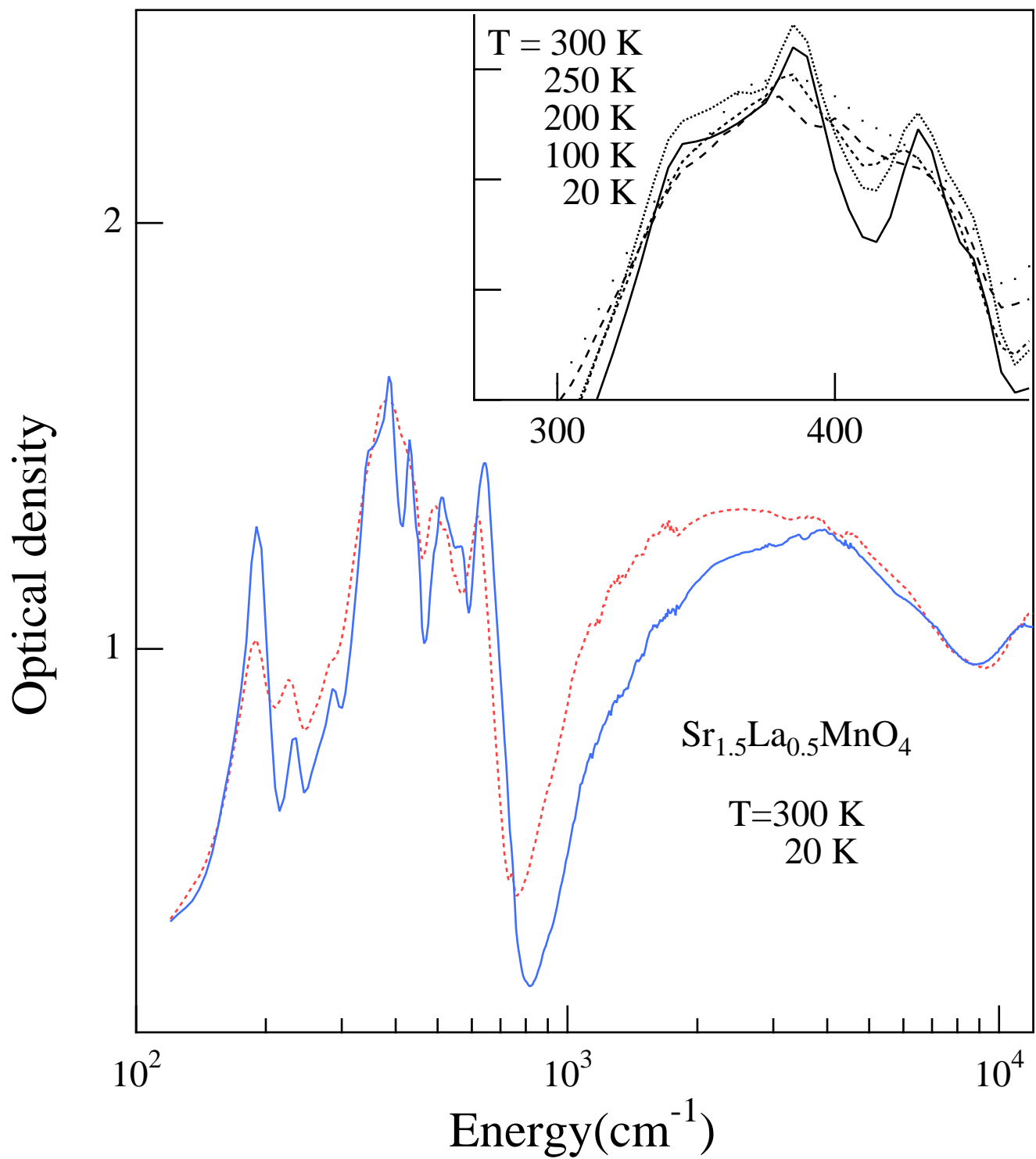
\end{document}